\documentclass[twocolumn]{aastex631}

\usepackage{url}
\usepackage{threeparttable}
\usepackage{ulem}
\usepackage{natbib}
\usepackage{appendix}
\usepackage{amsmath}
\usepackage{amssymb}
\usepackage{multirow}
\usepackage{float}
\usepackage{physics}

\def\ion#1#2{#1$\;${\sc\@roman{#2}}\relax}
\def\lesssim{\mathrel{\hbox{\rlap{\hbox{\lower4pt\hbox{$\sim$}}}\hbox{$<$}}}}
\def\gtrsim{\mathrel{\hbox{\rlap{\hbox{\lower4pt\hbox{$\sim$}}}\hbox{$>$}}}}
\def\red{\textcolor{black}}

\shorttitle{
[$\alpha$/Fe] in a Luminous Galaxy at $z>10$
}
\shortauthors{Nakane et al.}
\begin{document}
\title{
Low [O/Fe] Ratio in a Luminous Galaxy at the Early Cosmic Epoch ($z>10$):\\
Signature of Short Delay Time or Bright Hypernovae/Pair-Instability Supernovae?
%Low [O/Fe] Ratio in GN-z11 at the Early Cosmic Epoch ($z=10.60$):\\
%Signature of Short Delay Time or Bright Hypernovae/Pair-Instability Supernovae?
%Confirming Globular-Cluster Formation with Fe to CNO elements and\\
%Extremely Low $\alpha$/Fe Abundance Ratios of GN-z11 at $z=10.60$: \\ Connection with Globular Cluster Formation \\ and Fe Enrichment by Hypernovae/Pair-instability Supernovae
}

\author[0009-0000-1999-5472]{Minami Nakane}
\affiliation{Institute for Cosmic Ray Research, The University of Tokyo, 5-1-5 Kashiwanoha, Kashiwa, Chiba 277-8582, Japan}
\affiliation{Department of Physics, Graduate School of Science, The University of Tokyo, 7-3-1 Hongo, Bunkyo, Tokyo 113-0033, Japan}

\author[0000-0002-1049-6658]{Masami Ouchi}
\affiliation{National Astronomical Observatory of Japan, 2-21-1 Osawa, Mitaka, Tokyo 181-8588, Japan}
\affiliation{Institute for Cosmic Ray Research, The University of Tokyo, 5-1-5 Kashiwanoha, Kashiwa, Chiba 277-8582, Japan}
\affiliation{Department of Astronomical Science, SOKENDAI (The Graduate University for Advanced Studies), Osawa 2-21-1, Mitaka, Tokyo, 181-8588, Japan}
\affiliation{Kavli Institute for the Physics and Mathematics of the Universe (WPI), University of Tokyo, Kashiwa, Chiba 277-8583, Japan}

\author[0000-0003-2965-5070]{Kimihiko Nakajima}
\affiliation{National Astronomical Observatory of Japan, 2-21-1 Osawa, Mitaka, Tokyo 181-8588, Japan}

\author[0000-0002-6047-430X]{Yuichi Harikane} 
\affiliation{Institute for Cosmic Ray Research, The University of Tokyo, 5-1-5 Kashiwanoha, Kashiwa, Chiba 277-8582, Japan}

\author[0000-0001-8537-3153]{Nozomu Tominaga}
\affiliation{National Astronomical Observatory of Japan, 2-21-1 Osawa, Mitaka, Tokyo 181-8588, Japan}
\affiliation{Astronomical Science Program, Graduate Institute for Advanced Studies, SOKENDAI, 2-21-1 Osawa, Mitaka, Tokyo 181-8588, Japan}
\affiliation{Department of Physics, Faculty of Science and Engineering, Konan University, 8-9-1 Okamoto, Kobe, Hyogo 658-8501, Japan}

\author[0000-0002-6705-6303]{Koh Takahashi}
\affiliation{National Astronomical Observatory of Japan, 2-21-1 Osawa, Mitaka, Tokyo 181-8588, Japan}

\author[0000-0001-9044-1747]{Daichi Kashino}
\affiliation{National Astronomical Observatory of Japan, 2-21-1 Osawa, Mitaka, Tokyo 181-8588, Japan}

\author[0000-0001-8057-4802]{Hiroto Yanagisawa}
\affiliation{Institute for Cosmic Ray Research, The University of Tokyo, 5-1-5 Kashiwanoha, Kashiwa, Chiba 277-8582, Japan}
\affiliation{Department of Physics, Graduate School of Science, The University of Tokyo, 7-3-1 Hongo, Bunkyo, Tokyo 113-0033, Japan}

\author[0000-0002-2740-3403]{Kuria Watanabe}
\affiliation{Department of Astronomical Science, SOKENDAI (The Graduate University for Advanced Studies), Osawa 2-21-1, Mitaka, Tokyo, 181-8588, Japan}
\affiliation{National Astronomical Observatory of Japan, 2-21-1 Osawa, Mitaka, Tokyo 181-8588, Japan}

\author[0000-0001-9553-0685]{Ken'ichi Nomoto}
\affiliation{Kavli Institute for the Physics and Mathematics of the Universe (WPI), University of Tokyo, Kashiwa, Chiba 277-8583, Japan}

\author[0000-0001-7730-8634]{Yuki Isobe}
\affiliation{Waseda Research Institute for Science and Engineering, Faculty of Science and Engineering, Waseda University, 3-4-1, Okubo, Shinjuku, Tokyo 169-8555, Japan}

\author[0000-0003-4321-0975]{Moka Nishigaki}
\affiliation{Department of Astronomical Science, SOKENDAI (The Graduate University for Advanced Studies), Osawa 2-21-1, Mitaka, Tokyo, 181-8588, Japan}
\affiliation{National Astronomical Observatory of Japan, 2-21-1 Osawa, Mitaka, Tokyo 181-8588, Japan}

\author[0000-0003-4656-0241]{Miho N. Ishigaki}
\affiliation{National Astronomical Observatory of Japan, 2-21-1 Osawa, Mitaka, Tokyo 181-8588, Japan}

\author[0000-0001-9011-7605]{Yoshiaki Ono}
\affiliation{Institute for Cosmic Ray Research, The University of Tokyo, 5-1-5 Kashiwanoha, Kashiwa, Chiba 277-8582, Japan}

\author[0009-0005-2897-002X]{Yui Takeda}
\affiliation{Department of Astronomical Science, SOKENDAI (The Graduate University for Advanced Studies), Osawa 2-21-1, Mitaka, Tokyo, 181-8588, Japan}
\affiliation{National Astronomical Observatory of Japan, 2-21-1 Osawa, Mitaka, Tokyo 181-8588, Japan}



\begin{abstract}
%250 words
We present an [O/Fe] ratio of a luminous galaxy GN-z11 at $z=10.60$ derived with the deep public JWST/NIRSpec data. We fit the medium-resolution grating \red{(G140M, G235M, and G395M)} data with the model spectra consisting of BPASS-stellar and \textsc{Cloudy}-nebular spectra in the rest-frame UV wavelength ranges with Fe absorption lines, carefully masking out the other emission and absorption lines in the same manner as previous studies conducted for lower redshift ($z\sim 2-6$) galaxies with oxygen abundance measurements. We obtain an Fe-rich abundance ratio \red{$\mathrm{[O/Fe]}=-0.37^{+0.43}_{-0.22}$, which} is confirmed with the independent deep prism data as well as by the classic 1978 index method. This [O/Fe] measurement is lower than \red{measured for star-forming galaxies} at $z\sim 2-3$. Because $z=10.60$ is an early epoch after the Big Bang ($\sim 430$ Myr) and first star formation (likely $\sim 200$ Myr), \red{it is difficult to produce} Fe by Type Ia supernovae (SNeIa) requiring sufficient delay time for white-dwarf formation and gas accretion. The Fe-rich abundance ratio in GN-z11 suggests that the delay time is short, or that the major Fe enrichment is not accomplished by SNeIa but bright hypernovae (BrHNe) and/or pair-instability supernovae (PISNe), where the yield models of BrHNe and PISNe explain Fe, Ne, and O abundance ratios of GN-z11. The [O/Fe] measurement is not too low to rule out the connection between GN-z11 and globular clusters (GCs) previously suggested by the nitrogen abundance, but rather \red{supports} the connection with a GC population at high [N/O] if a metal dilution process exists.

\end{abstract}


%% subsec tion for cross-referencing in a \ref command.

\section{Introduction} \label{sec:introduction}

The chemical abundance ratios of galaxies provide insights into stellar nucleosynthesis. Of particular interest is $\alpha$-element abundances relative to iron, $\alpha$/Fe (e.g., O/Fe), which trace the star formation timescale. The $\alpha$-elements are almost instantaneously ejected by core-collapse supernovae (CCSNe) at the end of the evolution of massive stars \red{(e.g., \citealt{Nomoto2013})} while a large fraction of iron is ejected by Type-Ia SNe (SNeIa), which require delay time for the formation of white dwarfs and gas accretion \red{(e.g., \citealt{Nomoto2018})}. 
\red{The abundance measurements for Milky Way (MW) stars (e.g., \citealt{Bensby2014}) and chemical evolution models with the ejecta of CCSNe and SNeIa (e.g., \citealt{Suzuki2018}) suggest that the $\alpha$/Fe ratios are higher than solar value due to CCSNe at low metallicity, and then begin to decrease when SNeIa occur and eject a large amount of iron.}
%At low metallicity, the $\alpha$/Fe ratios are higher than solar value due to CCSNe. When SNeIa occur and eject large amount of iron, the $\alpha$/Fe ratios begin to decrease. This trend is observed for Milky Way (MW) stars (e.g., \citealt{Bensby2014}). 
However, before the appearance of SNeIa, low-$\alpha$/Fe gas can be produced by hypernovae (HNe) and theoretical pair-instability SNe (PISNe). These SNe have more than $10$ times larger explosion energy than standard CCSNe (\citealt{Heger&Woosley2002,Umeda&Nomoto2002,Nomoto2004,Nomoto2006,Takahashi2016}), causing massive core destruction and ejecting a lot of iron. \citet{Umeda&Nomoto2002} predict that HNe with higher explosion energies are likely to produce lower O/Fe gas below the solar abundance ratio. Hereafter, we refer to such HNe with high explosion energies as bright HNe (BrHNe). 
\begin{figure*}[ht!]
    \centering
    \includegraphics[scale=0.55]{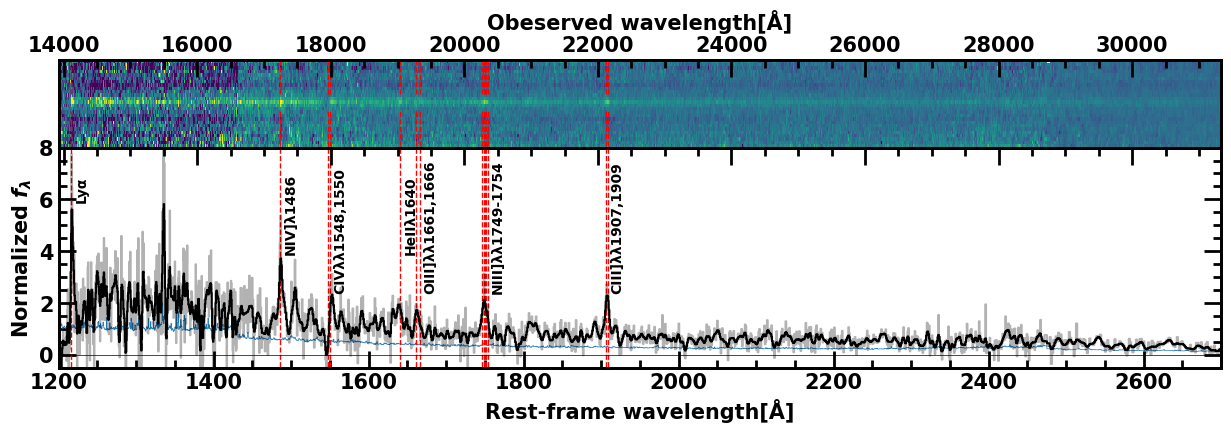}
    \includegraphics[scale=0.542]{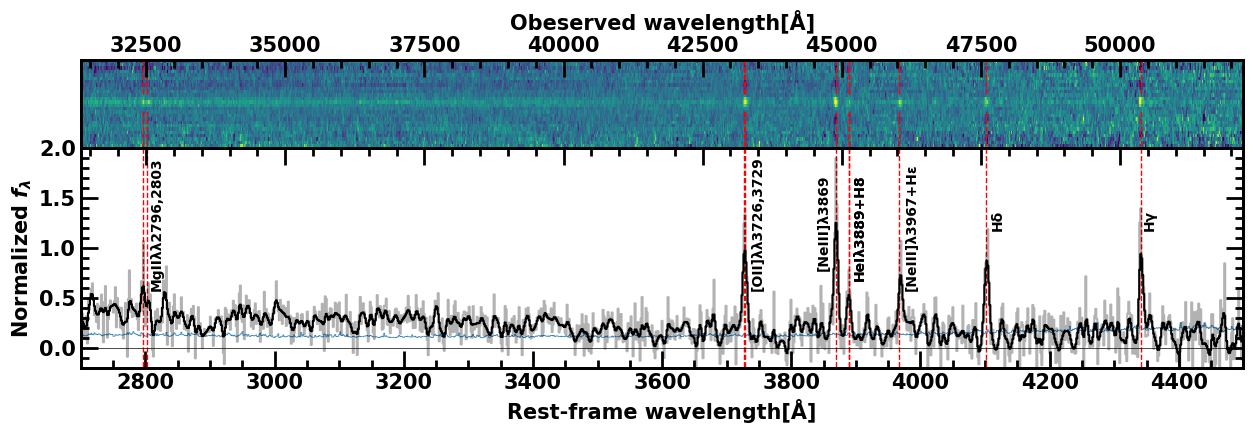}
    \caption{Stacked grating spectrum of GN-z11. Each panel shows the two-dimensional spectrum (\red{upper panel}) and one-dimensional spectra (\red{lower panel}). The gray and blue solid lines indicate the observed spectrum and its $1\sigma$ error spectrum, respectively. The black line represents the spectrum smoothed with an arbitrary Gaussian kernel of $\sigma=1.5$ \AA. The red-dashed lines denote emission lines detected in \citet{Bunker2023b}.}
    \label{fig:stack}
\end{figure*}

Abundance ratios of O/Fe have been measured for galaxies. \red{In the local universe, the O/Fe ratios in extremely metal-poor galaxies (EMPGs) are measured from emission line ratios (e.g., \citealt{Kojima2021,Isobe2022,Watanabe2024}). Some EMPGs have unusually low O/Fe ratios, which is not explained by the ejecta of CCSNe only, but reproduced by adding the ejecta of BrHNe and/or PISNe \citep{Isobe2022}.} Comparing the Ar/O and S/O ratios of the Fe-rich EMPGs with the chemical evolution models, \citet{Watanabe2024} suggest that the abundance ratios of the Fe-rich EMPGs are not explained by PISNe, but by the combination of \red{CCSNe/HNe and SNeIa with the short delay time.} 
Such extreme supernovae like HNe, BrHNe, and PISNe are not found even in the high-redshift universe. Although it is difficult to measure iron \red{abundances} at high redshift due to the weak iron emission lines, stellar photospheric absorption lines in the rest-UV continuum dominated mainly by iron-peak element \citep{Steidel2016} enable us to obtain iron \red{abundances}. Some studies derive O/Fe for galaxies at $z\sim2-3$ and $6$ by comparing observed UV composite spectra with synthetic spectra from stellar population models \citep{Steidel2016,Cullen2019,Harikane2020,Cullen2021,Kashino2022}. They claim that the O/Fe value of the galaxies at $z\sim2-3$ and $6$ is enhanced compared to the solar value \red{seen in $z\sim0$ galaxies, due to their higher fraction of recently-formed stars}. The similarity of high O/Fe between high-redshift galaxies \red{and} the old stars in Galactic thick disk is also pointed out \citep{Steidel2016,Cullen2021,Kashino2022}. 

Recent studies report low O/Fe ratios for GS\_3073 ($z=5.55$; \citealt{Ji2024a}) and GS9422 ($z=5.943$; \citealt{Tacchella2024}) with the Fe emission lines based on the observations of the James Webb Space Telescope (JWST; \citealt{Ferruit2022,Jakobsen2022,Boker2023,Rieke2023,Rigby2023}). \citet{Ji2024a} and \citet{Tacchella2024} suggest that the Fe enrichment is caused by SNeIa and/or HNe. To distinguish between SNeIa and HNe/PISNe as the origin of low O/Fe ratio, it is important to observe the early universe, when SNeIa have not occurred due to the delay time.

In this study, we measure the O/Fe ratio of GN-z11, a luminous ($M_\mathrm{UV}=-21.5$) galaxy at $z=10.60$ (e.g., \citealt{Bunker2023a}), from the spectra obtained with JWST/NIRSpec. Comparing O/Fe and other abundance ratios of GN-z11 with lower redshift galaxies, Milky Way stars, and models, we discuss the star formation in GN-z11. This paper is organized as follows. Section \ref{sec:data} \red{describes} the JWST/NIRSpec observations and spectroscopic data of GN-z11. We measure the stellar metallicity from spectral fitting and derive abundance ratios of O/Fe in Section \ref{sec:chemical}. In Section \ref{sec:discussion}, we discuss the origins of \red{the} observed O/Fe ratio and \red{a} relationship between the globular cluster (GC) stars and GN-z11. Section \ref{sec:summary} summarizes our findings. We assume a standard $\Lambda$CDM cosmology with $\Omega_\Lambda=0.7$, $\Omega_m=0.3$, and $H_0=70$ km $\mathrm{s}^{-1}$ $\mathrm{Mpc}^{-1}$. All magnitudes are in the AB system \citep{Oke&Gunn1983}. Throughout this paper, we use the solar abundance ratios of \citet{Asplund2021}. The notation [X/Y] is defined as log(X/Y) subtracted by the solar abundance ratio of log$(\mathrm{X/Y})_\odot$.

%\clearpage
\section{Data} \label{sec:data}

The spectroscopic data of GN-z11 were obtained in the Guaranteed Time Observations (GTO) of the JWST Advanced Deep Extragalactic Survey (JADES; GTO-1181, PI: D. Eisenstein; \citealt{Eisenstein2023}), using NIRSpec \citep{Jakobsen2022} in its microshutter array (MSA) mode \citep{Ferruit2022}. The observations of GN-z11 were conducted in two GTO-1181 programs of GOODS-N Medium/HST and Medium/JWST, where `Medium' and `HST' (`JWST') describe the depth and telescope used for target selection, respectively. Hereafter, we refer to the spectra of GN-z11 observed in Medium/HST and Medium/JWST as JADES-3991 and JADES-5591, respectively. JADES-3991 (JADES-5591) was \red{obtained} in February 2023 (May 2023) at 4 pointings with the prism (resolution $R\sim100$) and medium resolution ($R\sim1000$) filter-grating pairs of F070LP-G140M, F170LP-G235M, and F290LP-G395M covering the wavelength of $0.6-5.3$, $0.7-1.3$, $1.7-3.1$, and $2.9-5.1$ $\mu\mathrm{m}$, respectively. \red{The total exposure time of JADES-3991 is $6.9$ hours for the prism and $3.5$ hours for each filter-grating pair while that of JADES-5591 is $2.7$ hours for the prism and each filter-grating pair. In total, the exposure time of GN-z11 is $9.6$ and $6.2$ hours for prism and each filter-grating pair, respectively.} We use the JADES data which are publicly available\footnote{\url{https://jades-survey.github.io/scientists/data.html}}, and have already \red{been} reduced with the pipeline developed by the ESA NIRSpec Science Operations Team and the NIRSpec GTO Team \red{\citep{Alves2018,Ferruit2022}}. \red{In the data processing steps, the pipeline optimizes the background subtraction, rectification, 1D extraction, and spectra combination steps for the targets observed in JADES programs \citep{Bunker2023b,D'Eugenio2024}.} The redshift of GN-z11, $z=10.6034$, is \red{spectroscopically} determined by weighted fit of the $8$ emission lines with \red{the signal to noise ratio of} S/N$>5$ (e.g., [Ne \textsc{iii}] $\lambda3869$ and H$\gamma$) from the grating spectra in \citet{Bunker2023b}. See \citet{Bunker2023a} and \citet{D'Eugenio2024} for further details of the observations and data reduction. In the section below, we focus on stacked spectrum of GN-z11 constructed from JADES-3991 and JADES-5591. We take error weighted mean flux values to calculate the stacked grating/prism spectra of GN-z11 from JADES-3991 and JADES-5591. We shift the stacked spectra of GN-z11 into the rest-frame wavelength based on the spectroscopic redshift, $z=10.6034$. In Figure \ref{fig:stack}, we show the stacked grating spectrum of GN-z11 normalized by the median flux in the wavelength range of $1270-2100$ \AA. 

%\includegraphics[]{}
%\clearpage
\section{Chemical Abundances} \label{sec:chemical}
\subsection{Measurement of Stellar Metallicity } \label{subsec:metallicity}
\begin{figure*}[t]
    \centering
    \includegraphics[scale=0.55]{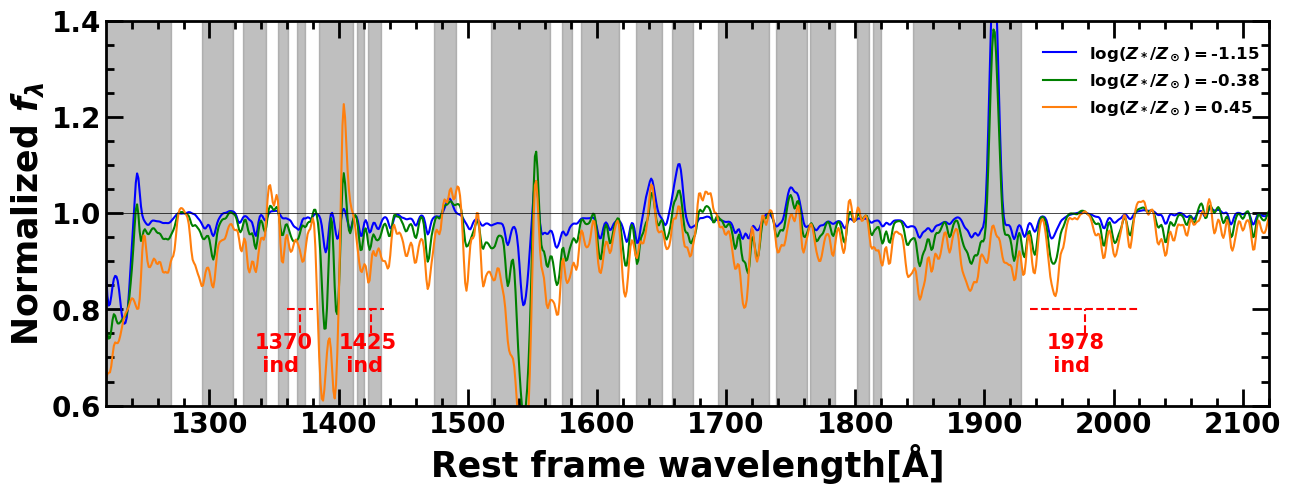}
    \caption{Examples of the model spectra. The blue, green, and orange lines present the model spectra with no dust attenuation, the age of $100$ Myr, and stellar metallicity of $\mathrm{log}(Z_*/Z_\odot)=-1.15$, $-0.38$, and $-0.45$ respectively. These model spectra are normalized by the pseudo continuum determined in the same way as \citet{Rix2004}. The gray-shaded regions are not used for the fitting (see Table \ref{tab:mask}). The red-dashed lines indicate \red{the wavelength ranges of the $1370$, $1425$, and $1978$} indices of the stellar metallicity suggested by \citet{Rix2004}.}
    \label{fig:model}
\end{figure*}
\begin{figure*}[t]
    \centering
    \includegraphics[scale=0.45]{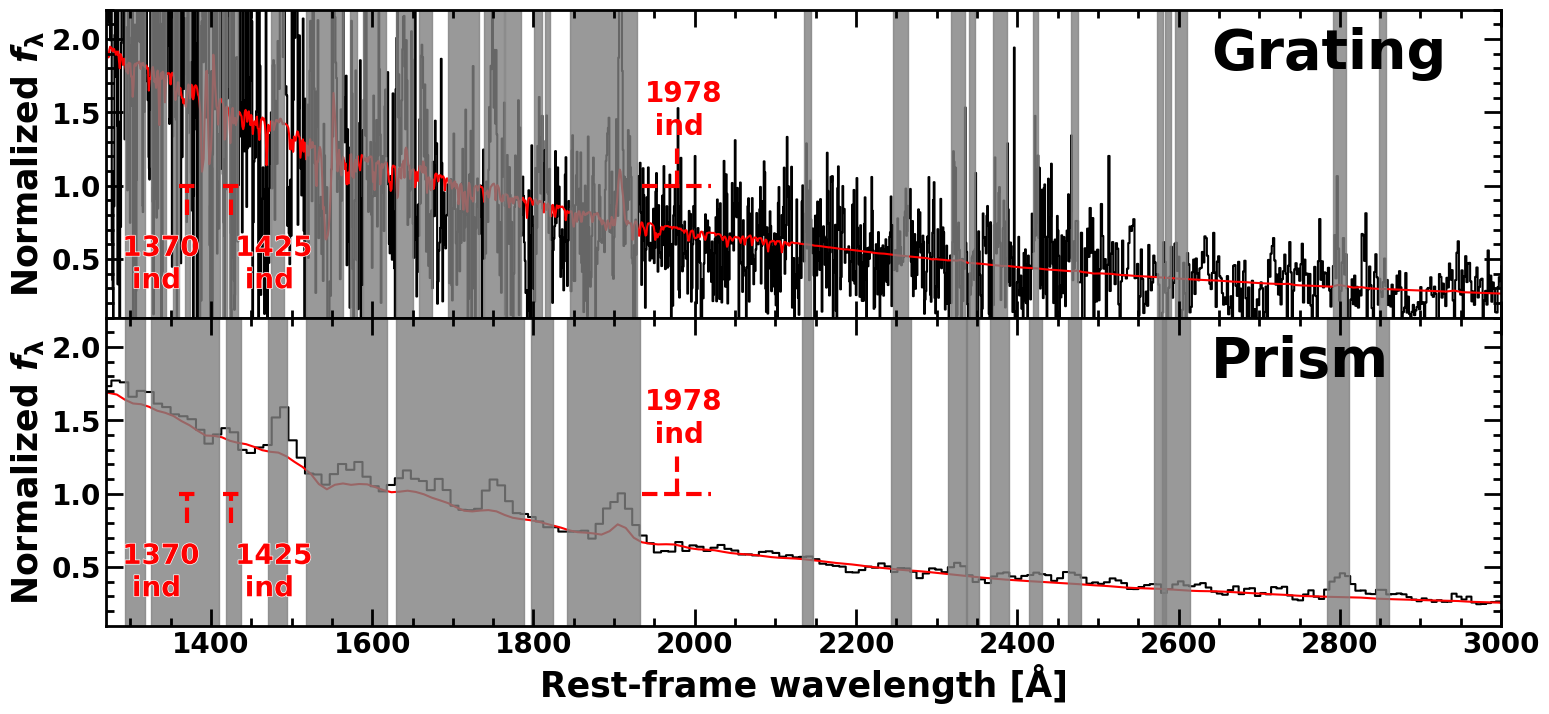}
    \caption{Results of the fitting. Top: the  black and red solid lines represent the stacked grating spectrum of GN-z11 and best-fit model spectrum, respectively. \red{The gray-shaded regions and red dashed lines show the same as Figure \ref{fig:model}}. Bottom: same as the top panel, but for the stacked prism spectra of GN-z11.}
    \label{fig:fit}
\end{figure*}

\begin{figure}
    \centering
    \includegraphics[scale=0.25]{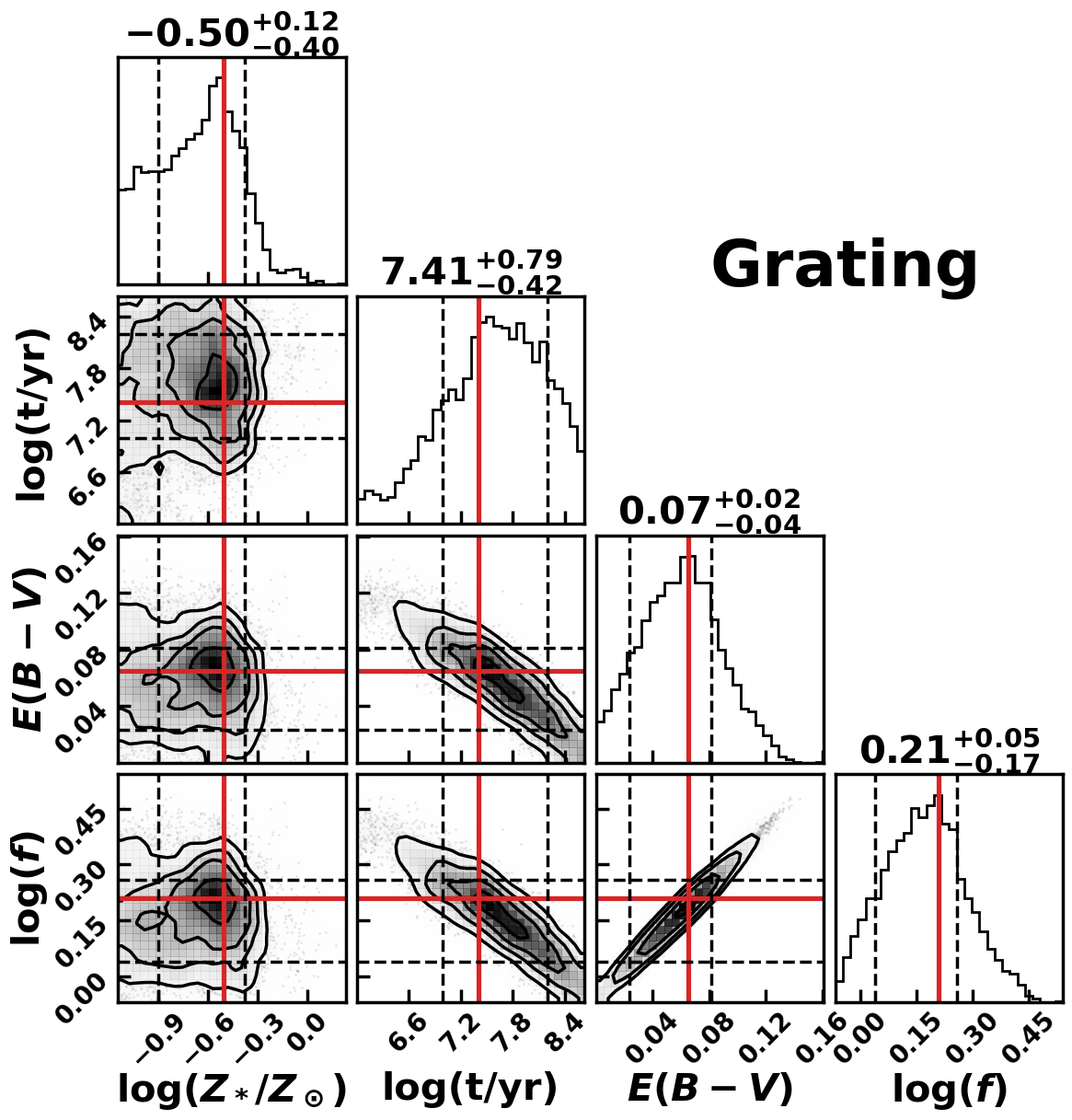}
    \includegraphics[scale=0.25]{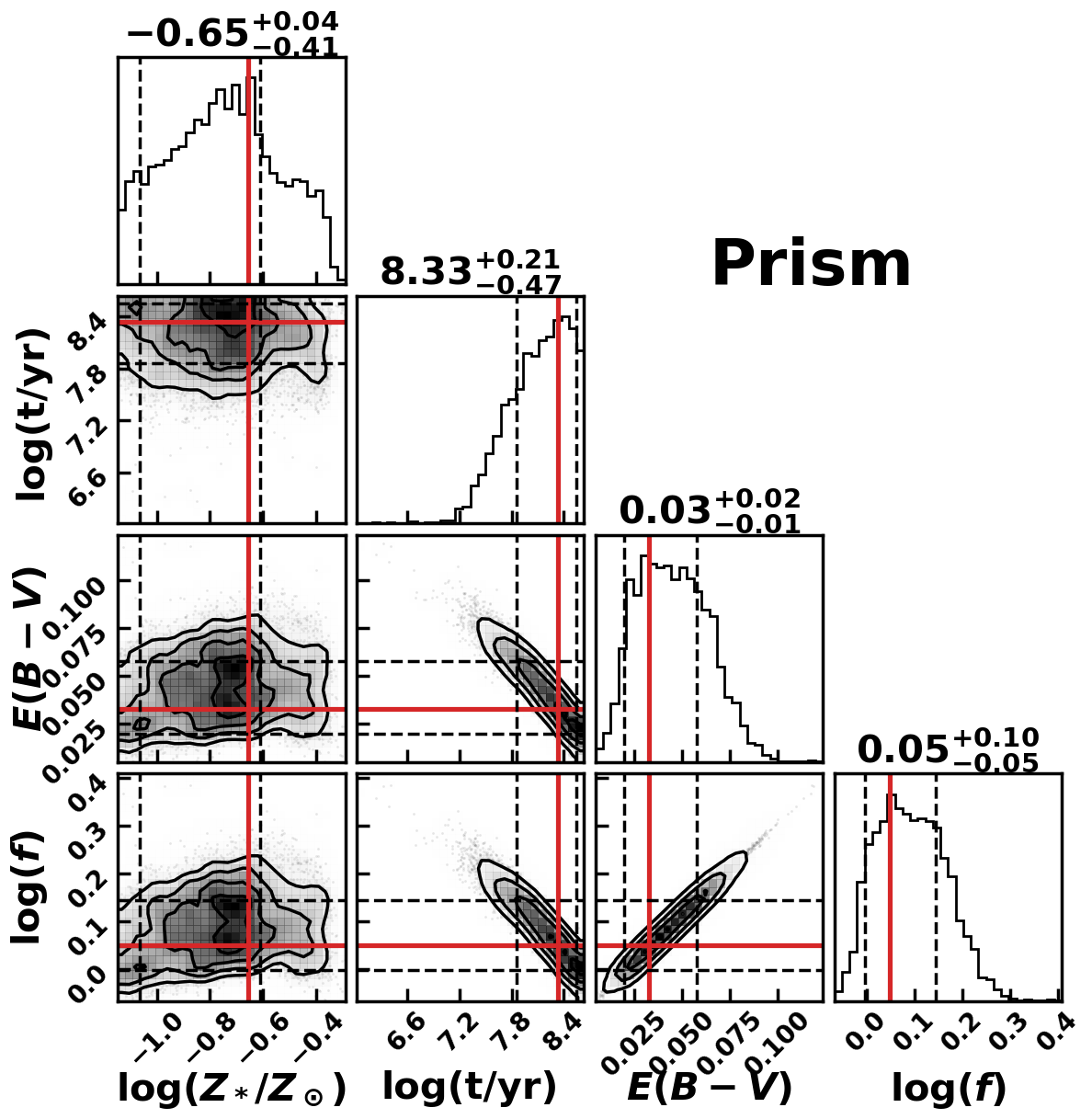}
    \caption{Posterior PDFs of the fitting parameters \red{obtained for the stacked grating (upper panels) and prism (lower panels) spectra of GN-z11.} The red solid lines and black-dashed lines represent the mode and the boundaries of the $68$th percentile HPDI, respectively (Section \ref{subsec:metallicity}).}
    \label{fig:pdf}
\end{figure}

To measure the stellar metallicity, we compare the observed spectra with model spectra which have different metallicities following the techniques described in previous studies \citep{Steidel2016,Cullen2019,Harikane2020,Kashino2022}. Here, we assume that the rest-frame UV continuum is dominated by the stellar radiation throughout our analysis (see Section \ref{subsec:abundance}). For our model spectra, we use the population synthesis code ``Binary Population and Spectral Synthesis" (BPASS v2.0; \citealt{Eldridge&Stanway2016,Stanway2016}) with different stellar metallicities ($Z_*=0.001$, $0.002$, $0.003$, $0.004$, $0.006$, $0.008$, $0.010$, $0.014$, $0.020$, $0.030$, $0.040$), adopting the \citet{Salpeter1955} IMF including binary stars and a continuous star formation history with different ages. The age $t$ is in the range from $10^{6}$ to $10^{10}$ yr incremented in steps of $0.1$ in log($t$/yr). Because the BPASS spectra do not include nebular emission, we add nebular continuum to the BPASS spectra using the photoionization code, \textsc{Cloudy} v23.01 \citep{Ferland1998,Gunasekera2023}. We calculate the nebular continuum, adopting the BPASS spectra as the incident spectrum. 
\red{We adopt a plane-parallel geometry, unity for a covering factor, the ionization parameter $\mathrm{log}(U)=-2.0$ of GN-z11 \citep{Bunker2023b}, and  \red{a} typical high-redshift value for electron density $n_e=300$ $\mathrm{cm^{-3}}$ \citep{Steidel2016}. We note that recent studies report high electron densities ($n_e\sim1000$ $\mathrm{cm^{-3}}$) measured from resolved [O\textsc{ii}] doublets for galaxies at $z\gtrsim8$ \citep{Isobe2023a,Marconcini2024,Abdurro'uf2024}. We confirm that high $n_e$ values does not affect our \red{conclusions}\footnote{\red{If we assume $n_e=1000$ $\mathrm{cm^{-3}}$, we obtain the stellar metallicity of $\log(Z_*/Z_\odot)=-0.53^{+0.41}_{-0.11}$, which is consistent with our fiducial value (Section \ref{subsec:metallicity}).}}.}
%We assume a plane-parallel geometry, unity for a covering factor, and typical high-redshift values for electron density $n_e=300$ $\mathrm{cm^{-3}}$ and ionization parameter $\mathrm{log}(U)=-2.8$ \citep{Steidel2016}. We confirm that changing $n_e$ and $\mathrm{log}(U)$ values does not affect our measurements of $Z_*$. 
The resulting spectra are reddened using \citet{Calzetti2000} dust extinction curve parameterized by a color excess in the range of $0.0\leq E(B-V)\leq1.0$, and attenuated by IGM absorption following the approach shown in \citet{Inoue2014}. The model spectra are smoothed to match the observed resolutions of $R=100$ and $1000$ for prism and grating spectra, respectively. \red{We do not take the wavelength dependence of the resolutions into account, which does not change our conclusions.} In Figure \ref{fig:model}, we show the model spectra normalized by the pseudo-continuum which we derive in the same way as \citet{Rix2004}. We confirm that the strengths of absorption lines from massive stars at $1360-1380$, $1415-1435$, and $1935-2020$ \AA\ ($1370$, $1425$, and $1978$ indices; \citealt{Rix2004}) significantly depend on the stellar metallicities.

We fit the model spectra to the stacked spectra of GN-z11 in the wavelength range of $1270-3000$ \AA. We first exclude the wavelength regions where the stellar continuum is impacted by nebular emission lines and interstellar absorption lines to derive stellar metallicity.
%Here, we note that the model spectra assume solar abundance ratios. We thus need to mask out non-iron photospheric absorption lines to convert log($Z_*/Z_\odot$) into [Fe/H]. 
\red{Because our model spectra assume a solar abundance pattern, we mask out wavelengths impacted by non-iron stellar absorption lines. By doing so, we make sure that the estimated log($Z_*/Z_\odot$) reflects the iron abundance [Fe/H] and is not biased by the non-solar abundance patterns (e.g., super-solar N/O; Section \ref{subsec:GC}.) actually reported in previous studies (\citealt{Cameron2023,Isobe2023b,Senchyna2024}).}
\begin{table*}[t]
    \centering
    \caption{Rest-frame Wavelength Ranges Excluded from Fitting \red{of the Grating Spectra}}
    \resizebox{\textwidth}{!}{
    \begin{tabular}{ccccc}
    \hline
    \hline
    $\lambda_{\mathrm{min}}$ (\AA)& $\lambda_{\mathrm{max}}$ (\AA)& Photospheric absorption&Interstellar absorption&Nebular emission\\
%    (1)&(2)&(3)&(4)&(5)\\
    \hline
    $1294$&$1318$&&O \textsc{i} $\lambda1302$, Si \textsc{iii} $\lambda\lambda1304,1309$&\\
    $1326$&$1344$&O \textsc{iv} $\lambda1341$&C \textsc{ii} $\lambda1334$&\\
    $1353$&$1361$&&O \textsc{i} $\lambda\lambda1356,1359$&\\
    $1368$&$1374$&O \textsc{v} $\lambda1371$&&\\
    $1385$&$1411$&&Si \textsc{iv} $\lambda\lambda1394,1402$&\\
    $1414$&$1420$&&Si \textsc{iv} $\lambda1417$&\\
    $1423$&$1433$&Ni \textsc{iv} $\lambda1426,1430$, C \textsc{iii} $\lambda1427$&&\\
    $1474$&$1491$&C \textsc{iii} $\lambda1478$&&N \textsc{iv}] $\lambda\lambda1483,1486$\\
    $1518$&$1564$&&Si \textsc{ii} $\lambda1526$, C \textsc{iv} $\lambda\lambda1548,1550$, C \textsc{i} $\lambda1560$&\\
    $1573$&$1581$&C \textsc{iii} $\lambda1577$&&\\
    $1588$&$1617$&C \textsc{iii} $\lambda1591$, O \textsc{v} $\lambda1596$&Fe \textsc{ii} $\lambda\lambda1608,1611$&\\
    $1630$&$1650$&&&He \textsc{ii} $\lambda1640$\\
    $1658$&$1674$&&Al \textsc{ii} $\lambda1670$&O \textsc{iii}] $\lambda\lambda1661,1666$\\
    $1694$&$1733$&N \textsc{iv} $\lambda\lambda\lambda1697,1699,1702$, N \textsc{iv} $\lambda1718$,&Ni \textsc{ii} $\lambda1710$&\\
    &&Si \textsc{iv} $\lambda\lambda1723,1727$, N \textsc{iii} $\lambda1730$&&\\
    $1739$&$1763$&&Ni \textsc{ii} $\lambda\lambda1742,1752$&N \textsc{iii}] $\lambda1750$(quintet), C \textsc{ii} $\lambda1761$\\
    $1765$&$1784$&O \textsc{iii} $\lambda\lambda\lambda1768,1773,1781$&&\\
    $1801$&$1811$&N \textsc{iii} $\lambda\lambda1804,1806$&Si \textsc{ii} $\lambda1808$&\\
    $1814$&$1820$&&&Si \textsc{ii} $\lambda\lambda1816,1817$\\
    $1845$&$1928$&O \textsc{iii} $\lambda\lambda\lambda\lambda1848,1857,1874,1921$, &Al \textsc{iii} $\lambda\lambda1855,1863$&Si \textsc{iii}] $\lambda\lambda1883,1892$, C \textsc{iii}] $\lambda\lambda1907,1909$\\
    &&N \textsc{iii} $\lambda1885$ C \textsc{iii} $\lambda\lambda1894,1923$&&\\
    $2136$&$2144$&&&N \textsc{ii}] $\lambda2140$\\
    $2246$&$2265$&&Fe \textsc{ii} $\lambda\lambda2250,2261$&\\
    $2318$&$2335$&&&[O \textsc{iii}] $\lambda\lambda2322,2331$, C \textsc{ii}] $\lambda2326$\\
    $2340$&$2348$&&Fe \textsc{ii} $\lambda$2344&\\
    $2370$&$2387$&&Fe \textsc{ii} $\lambda\lambda2374,2383$&\\
    $2420$&$2426$&&&[Ne \textsc{iv}] $\lambda2423$\\
    $2467$&$2475$&&&[O \textsc{ii}] $\lambda\lambda$2471\\
    $2573$&$2581$&&Mn \textsc{ii} $\lambda2577$&\\
    $2583$&$2591$&&Mn \textsc{ii} $\lambda2587$&\\
    $2596$&$2610$&&Fe \textsc{ii} $\lambda2600$, Mn \textsc{ii} $\lambda2600$&\\
    $2792$&$2808$&&Mg \textsc{ii} $\lambda\lambda2796,2803$&\\
    $2849$&$2857$&&Mg \textsc{i} $\lambda2853$&\\
%    $3185$&$3193$&&&He \textsc{i} $\lambda3189$\\
%    $3199$&$3207$&&&He \textsc{ii} $\lambda3203$\\
%    $3337$&$3351$&&&[Ne \textsc{iii}] $\lambda3342$, [Ne \textsc{v}] $\lambda3346$\\
%    $3422$&$3430$&&&[Ne \textsc{v} $\lambda3426$]\\
    \hline
    \end{tabular}}
    \label{tab:mask}
\end{table*}
As for the indices of \citet{Rix2004}, while the $1978$ index is mainly dominated by iron absorption lines of the Fe \textsc{iii} blend, the $1370$ ($1425$) index includes both iron and non-iron absorption lines of O \textsc{v} $\lambda1371$ and Fe \textsc{v} $\lambda\lambda1360-1380$ (Si \textsc{iii} $\lambda1417$, C \textsc{iii} $\lambda1427$, and Fe \textsc{v} $\lambda1430$). Therefore, we do not use the wavelength regions around non-iron photospheric absorption lines with reference to \citet{Dean&Bruhweiler1985}, \citet{Brandt1998}, and \citet{Leitherer2011}. The wavelength regions excluded from the fitting \red{of the grating spectra} are summarized in Table \ref{tab:mask}. \red{For the prism spectra, we mask out the wider wavelength ranges than shown in Table \ref{tab:mask} due to the low spectral resolution.} It is crucial to use the spectra in the wavelength range of $1270-3000$ \AA, especially the shorter wavelength, for the fitting to accurately measure the iron abundances. This is because the shorter wavelength ranges include the strong iron absorption lines like the $1370$ and $1425$ indices and are important to determine the overall continuum levels of the spectra. We do not use the spectra above $3000$ \AA\ because a continuum excess in $3000-3550$ \AA\ is reported (\citealt{Ji2024b}; see Section \ref{subsec:abundance}). Here, we confirm that if we change the wavelength range for the fitting to $1270-3500$ \AA\ (including the wavelength region of the reported continuum excess), our conclusion does not change\footnote{\red{The stellar metallicity measured from the fitting to the stacked grating spectrum of GN-z11 in the wavelength range of $1270-3500$ \AA\ is  log$(Z_*/Z_\odot)=-0.52^{+0.06}_{-0.46}$}.}\red{.} We estimate \red{four} free parameters\red{:} the stellar metallicity $Z_*$, age $t$, color excess $E(B-V)$, and normalization factor of the model spectra $f$. To obtain the posterior probability distribution functions (PDFs) of \red{these} parameters, we conduct Markov Chain Monte Carlo (MCMC) simulations, using \texttt{emcee} \citep{Foreman2013}. We use flat priors of $-1.15\leq\mathrm{log}(Z_*/Z_\odot)\leq0.45$, $6.00\leq\mathrm{log}(t/\mathrm{yr})\leq8.63$, $0.0\leq E(B-V)\leq1.0$, and $-2.0\leq\mathrm{log}(f)\leq2.0$. The ranges of $Z_*$ and $t$ are constrained by the possible choices in the BPASS models and cosmic age at $z=10.60$. Since the model spectra have only $11$ $(41)$ discrete metallicity (age) values, we linearly interpolate flux values in  $\mathrm{log}(Z_*)-\mathrm{log}(t)-\mathrm{log(flux)}$ space. We then obtain the model spectra at any metallicity within the prescribed range. We first conduct the fitting to the stacked grating spectrum of GN-z11. In the top panels of Figures \ref{fig:fit} and \ref{fig:pdf}, we show the best-fit model spectra and posterior PDFs of the free parameters, respectively. We determine the free parameter and $1\sigma$ uncertainty by the mode (i.e., a peak of the posterior distribution) and $68$th percentile highest posterior density interval (HPDI; i.e., the narrowest interval containing $68\%$) of the posterior distribution, respectively. The best-fit parameters for the stacked grating spectrum of GN-z11 are \red{log$(Z_*/Z_\odot)=-0.50^{+0.12}_{-0.40}$, log$(t/\mathrm{yr})=7.41^{+0.79}_{-0.42}$, $E(B-V)=0.07^{+0.02}_{-0.04}$, and log$(f)=0.21^{+0.05}_{-0.17}$.} As described above, we can \red{derive [Fe/H] by log$(Z_*/Z_\odot)=\mathrm{[Fe/H]}$, and then obtain $\mathrm{[Fe/H]}=-0.50^{+0.12}_{-0.40}$.} The best-fit $t$ and $E(B-V)$ are consistent with the young stellar age ($24^{+20}_{-10}$ Myr) and low dust attenuation ($A_V=0.08^{+0.23}_{-0.06}$ mag) derived from spectral energy distribution (SED) fitting for the photometry data of GN-z11 observed with JWST NIRCam \citep{Tacchella2023}. We conduct the same fitting to the stacked prism spectrum of GN-z11. We present the fitting results in the bottom panels of Figures \ref{fig:fit} and \ref{fig:pdf}. The measured stellar metallicity of \red{log$(Z_*/Z_\odot)=-0.65^{+0.04}_{-0.41}$} is consistent with that obtained from the stacked grating spectrum of GN-z11 within errors. 
\begin{figure}[t]
    \centering
    \includegraphics[scale=0.52]{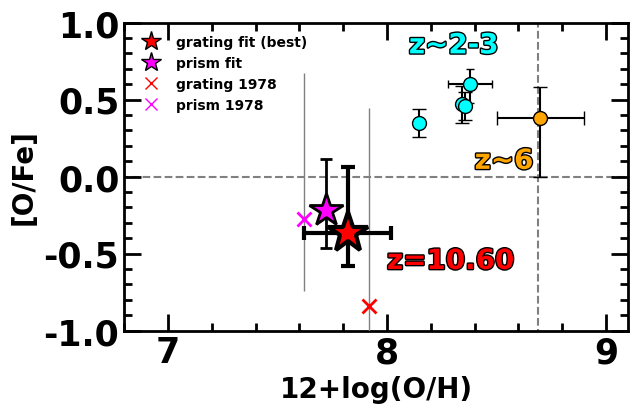}
    \caption{Comparison of O/Fe measurements based on the stellar photospheric absorption lines. The red and magenta star \red{symbols} indicate our measurements from \red{spectral fitting to the} stacked grating and prism spectra of GN-z11, respectively. The red and magenta crosses represent our measurements from the $1978$ index \citep{Rix2004} with \red{the} stacked grating and prism spectra of GN-z11, respectively. For \red{display purposes}, the magenta cross, magenta star \red{symbol}, and red cross are shifted by $-0.2$, $-0.1$, and $+0.1$ dexes in the x-axis, respectively. The cyan and orange circles show the measurements in the fitting method for composite spectra of galaxies at $z\sim2-3$ \citep{Steidel2016,Cullen2021,Kashino2022} and $z\sim6$ \citep{Harikane2020}, respectively. The gray-dashed lines denote the solar abundance ratios.}
    \label{fig:comparison}
\end{figure}
\begin{figure}[t]
    \centering
    \includegraphics[scale=0.55]{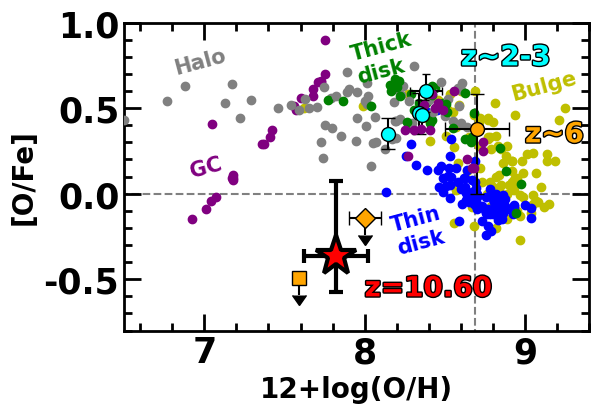}
    \caption{\red{Comparison with low-redshift galaxies and Galactic stars in the O/Fe vs. O/H diagram.} The red star \red{symbol}, cyan and orange circles, and gray-dashed lines represent the same as Figure \ref{fig:comparison}. The orange diamond and square show the results for GS\_3073 at $z=5.55$ \citep{Ubler2023,Ji2024a} and GS9422 at $z=5.943$ \citep{Cameron2024,Tacchella2024,Terp2024}, respectively. The blue, \red{green, yellow}, purple, and gray circles denote the results for MW stars \citep{Melendez2003,Carretta2005,Yong2005,Lecureur2007,Pasquini2008,Yong2008,Carretta2010,Valenti2011,Bensby2013,Zhao2016,Amarsi2019} in the thin disk, thick disk, bulge, globular cluster, and halo, respectively.}
    \label{fig:O_H-O_Fe}
\end{figure}
\begin{figure}[t]
    \centering
    \includegraphics[scale=0.54]{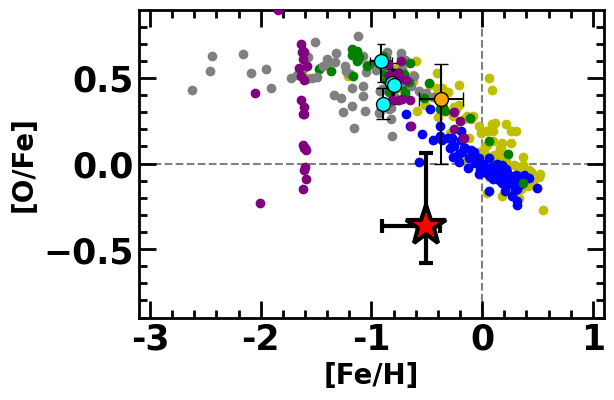}
    \caption{\red{Comparison with low-redshift galaxies and Galactic stars in the O/Fe vs. Fe/H diagram. Symbols are the same as in Figure \ref{fig:O_H-O_Fe}.}}
    \label{fig:Fe_H-O_Fe}
\end{figure}

For comparison to the stellar metallicity derived from a different method, we measure the $1978$ index of \citet{Rix2004}, that is, the equivalent width across $1935-2020$ \AA\ (EW($1978$)). We do not measure the other $1370$ and $1425$ indices because these indices are partly masked out due to the non-iron absorption lines described above. We determine the continuum fluxes of the stacked grating and prism spectra of GN-z11 with power-law fitting. We then calculate EW($1978$) by summing up the observed fluxes relative to the continuum levels which are divided by the continuum fluxes in the wavelength range of $1935-2020$ \AA. For \red{the} stacked prism (grating) spectrum of GN-z11, we obtain EW$(1978)=3.7^{+1.2}_{-1.9}$ \AA\ ($7.5^{+2.5}_{-4.8}$ \AA), which is converted into log$(Z_*/Z_\odot)=-0.59^{+0.27}_{-0.75}$ ($-0.03^{+1.69}_{-1.09}$) by using Equation (8) in \citet{Rix2004}. The $Z_*$ values measured with the $1978$ index are consistent with those from the fitting method within errors (see Section \ref{subsec:abundance}). We note that the uncertainty is larger than those from the fitting method because the wavelength regions used for the $1978$ index are limited.

%\clearpage
\subsection{Abundance Ratios of $\alpha$/Fe} \label{subsec:abundance}

To derive O/Fe for GN-z11, we take the O/H value of 12+log(O/H)$=7.82\pm0.20$ derived \red{from multiple emission lines, assuming photoionization by stellar radiation \citep{Cameron2023}.} See \citet{Cameron2023} for further details of the method to measure the O/H value. \red{While the O/H value traces the metallicity from the ionized gas near the short-lived massive stars (e.g., \citealt{Steidel2016}), our Fe/H value reflects the metallicities of massive stars (Section \ref{subsec:metallicity}). Here, note that chemical abundances of massive stars should be similar to those of the surrounding ionized gas because their lifetimes are short  \citep{Steidel2016,Cullen2019,Harikane2020,Cullen2021,Kashino2022}. We thus adopt the ratio of the gas-phase O/H to the stellar Fe/H as a proxy \red{of the instantaneous} O/Fe in the galaxy.}
\begin{figure}[t]
    \centering
    \includegraphics[scale=0.55]{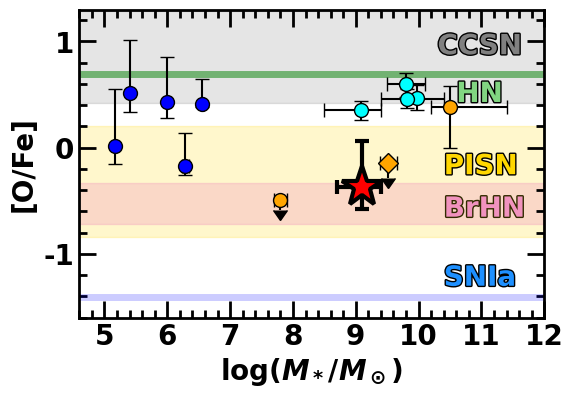}
    \caption{Abundance ratio of O/Fe as a function of stellar mass for galaxies. \red{Symbols are} the same as \red{in} Figure \ref{fig:O_H-O_Fe}. The blue circles indicate the measurements for local EMPGs \citep{Izotov2018,Kojima2020,Isobe2022}. The gray, green, violet, yellow, and blue shaded regions represent the yields of CCSNe, HNe \citep{Nomoto2013}, BrHNe \citep{Umeda&Nomoto2002}, PISNe \citep{Takahashi2018}, and SNeIa \citep{Nomoto1984,Iwamoto1999}, respectively.}
    \label{fig:O_Fe-M}
\end{figure}
\begin{figure}[t]
    \centering
    \includegraphics[scale=0.55]{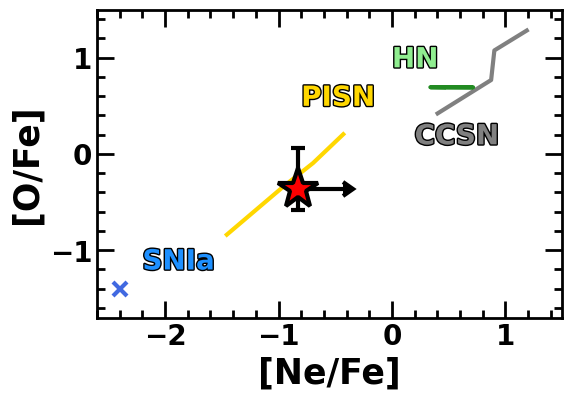}
    \caption{Comparison of O/Fe vs. Ne/Fe for GN-z11 and supernovae yields. The marker and lines represent the same as Figure \ref{fig:O_Fe-M}.}
    \label{fig:Ne_Fe-O_Fe}
\end{figure}
\begin{figure}[t]
    \centering
    \includegraphics[scale=0.58]{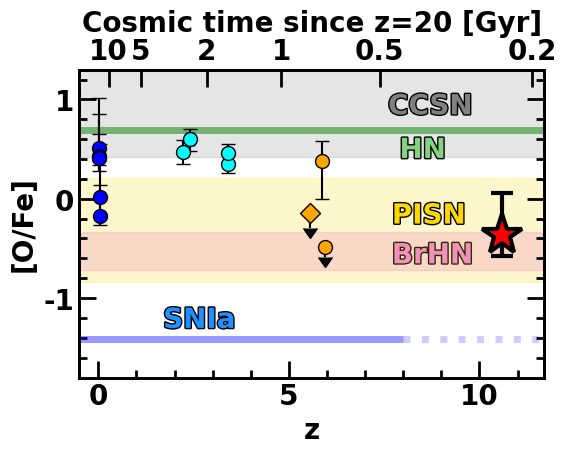}
    \caption{\red{Abundance ratio of O/Fe as a function of redshift. Symbols are the same as in Figure \ref{fig:O_Fe-M}.}}
    \label{fig:z-O_Fe}
\end{figure} 
In Figure \ref{fig:comparison}, we \red{present} our measurements of O/Fe from the stellar metallicities derived \red{with} the fitting and $1978$ index methods in Section \ref{subsec:metallicity}. Although the uncertainties of O/Fe values based on the $1978$ index are large, all measurement are consistent within errors. We thus use the stellar metallicity based on the stacked grating spectrum of GN-z11, log$(Z_*/Z_\odot)=\mathrm{[Fe/H]}=\red{-0.50^{+0.12}_{-0.40}}$, as the fiducial value. We then obtain $\mathrm{[O/Fe]}=\red{-0.37^{+0.43}_{-0.22}}$. In Figures \ref{fig:O_H-O_Fe} and \ref{fig:Fe_H-O_Fe}, we compare O/Fe of GN-z11 with those of MW stars \citep{Melendez2003,Carretta2005,Yong2005,Lecureur2007,Pasquini2008,Yong2008,Carretta2010,Valenti2011,Bensby2013,Zhao2016,Amarsi2019}, GS\_3073 ($z=5.55$; \citealt{Ubler2023,Ji2024a}), GS9422 ($z=5.943$; \citealt{Cameron2024,Tacchella2024,Terp2024}) and \red{star-forming} galaxies at $z\sim2-3$ \citep{Steidel2016,Cullen2021,Kashino2022} and $6$ \citep{Harikane2020}. We find that the O/Fe value of GN-z11 is lower than those of MW stars and lower redshift galaxies at the same O/H and Fe/H values. Such a rich Fe abundance of GN-z11 is also pointed out by \citet{Ji2024b}. In the spectrum of GN-z11, \citet{Ji2024b} report a continuum excess in the rest-frame $3000-3550$ \AA, which is associated with the Fe \textsc{ii} complex emission from the broad-line region of the active galactic nucleus (AGN). \citet{Ji2024b} claim the rich Fe abundance of $\mathrm{[Fe/H]}\gtrsim-0.5$ based on the AGN models (the combination of UV continuum, Balmer continuum, and Fe \textsc{ii} emission). We find that our measurement of Fe/H ($\mathrm{[Fe/H]}=\red{-0.50^{+0.12}_{-0.40}}$) based on the stellar models is consistent with the results based on the AGN models. In other words, GN-z11 is Fe rich population regardless of the origin of UV emission dominated by stars or AGNs.

We also focus on the other $\alpha$-element abundances of Ne and Mg, of which emission lines [Ne \textsc{iii}] $\lambda3869$ and Mg \textsc{ii} $\lambda\lambda2796,2803$ are detected in the grating spectrum (see Figure \ref{fig:stack}). 
For the Ne abundance, \citet{Isobe2023b} obtain the abundance ratio of $\mathrm{[Ne/O]}>-0.25$ from the [Ne \textsc{iii}] $\lambda3869$/[O \textsc{iii}] $\lambda4363$ ratio assuming the stellar radiation. Combining O/Fe measured in this work, we obtain $\mathrm{[Ne/Fe]}>-0.97$. 
\red{The Mg abundance can be estimated from the flux ratio of Mg\textsc{ii} $\lambda\lambda2796,2803$ and H$\beta$, assuming $\mathrm{Mg/H}=\mathrm{Mg^+}/\mathrm{H^+}\times \mathrm{ICF(Mg^+)}$, where $\mathrm{ICF(Mg^+)}$ is the ionization correction factor (ICF) to add the abundance of $\mathrm{Mg}^{2+}$ \citep{Guseva2013}. However, the $\mathrm{ICF(Mg^+)}$ value} depend on $\mathrm{O^{2+}/O}$ and largely change for \red{the} large value of $\mathrm{O^{2+}/O}\sim1$ (see Figure 4 in \citealt{Guseva2013}). Because GN-z11 have large value of $\mathrm{O^{2+}/O}\sim0.95$ \citep{Cameron2023}, the $\mathrm{ICF(Mg^+)}$ value is highly uncertain ($\sim15-70$). We are thus not able to constrain the Mg abundance for GN-z11.

\section{Discussion} \label{sec:discussion}
\subsection{Origin of Low O/Fe} \label{subsec:origin}

As described in Section \ref{subsec:abundance}, we find the low O/Fe ratio of GN-z11 compared to the MW stars and lower \red{redshift} galaxies (Figures \ref{fig:O_H-O_Fe} and \ref{fig:Fe_H-O_Fe}). Here, we explore the origins of such low O/Fe. As described in Section \ref{sec:introduction}, Fe enrichment is caused by SNeIa, HNe, BrHNe, and PISNe. In Figure \ref{fig:O_Fe-M}, we plot O/Fe as a function of stellar mass $M_*$ for GN-z11, GS\_3073, GS9422, local EMPGs, and galaxies at $z\sim2-3$ and $6$ with the yields of CCSNe, SNeIa, HNe, BrHNe, and PISNe for comparison. We take the $M_*$ value of GN-z11, $\mathrm{log}(M_*/M_\odot)=9.1^{+0.3}_{-0.4}$, from \citet{Tacchella2023}. The O/Fe values of GN-z11, GS\_3073, GS9422, and $z\sim6$ galaxies are lower than $z\sim2-3$ galaxies at the same $M_*$ value. 

Here, we briefly explain the yield models shown in Figure \ref{fig:O_Fe-M}. For the CCSNe and HNe models, we use the yields of \citet{Nomoto2013} with the progenitor mass of $20-40\ M_\odot$ and the metallicity of $Z_*=0.004$ which is comparable to our measurements. The CCSNe and HNe models are different in the explosion energy $E$. While the explosion energy is $E_{51}=E/10^{51}\ \mathrm{erg}=1$ for all of the CCSNe models, HNe models have larger explosion energies of $E_{51}=10$, $10$, $20$, and $30$ for the progenitor mass of $20$, $25$, $30$, and $40$ $M_\odot$, respectively. The Fe \red{abundance depends not only on} the explosion energy, but also \red{on the} mass cut that divides the ejecta and compact remnant. In HNe models, the mass cuts are chosen to explain the Fe ejecta constrained from the observed HNe (e.g., \citealt{Nomoto2004}). For BrHNe models, we use the yields of \citet{Umeda2008} with the highest explosion energy \red{of $E_{51}=50$, $150$, $100$, $110$, and $210$ for the progenitor mass of $30$, $50$, $80$, $90$ and $100$ $M_\odot$, respectively,} and lowest mass cut just above the Fe core for a certain progenitor mass \red{within} $30-100$ $M_\odot$ (see Table 3 in \citealt{Umeda2008}). Therefore, BrHNe eject the lowest O/Fe ratio gas among the HNe at the same progenitor mass. In the PISNe models, we take the nonrotating PISNe yields of \citet{Takahashi2018} with the progenitor mass of $220-280$ $M_\odot$. We use the SNeIa yields of \citet{Nomoto1984} and \citet{Iwamoto1999}. 

In Figure \ref{fig:O_Fe-M}, the O/Fe value of GN-z11 is lower than the yields of CCSNe and observed HNe, but comparable to the yields of BrHNe and PISNe. We note that the differences in the HNe and BrHNe models are the mass cut and explosion energy. As shown in Figure \ref{fig:Ne_Fe-O_Fe}, we confirm that the Ne/Fe value of GN-z11 \red{is} also comparable to the yields of PISNe. We do not present the yields of BrHNe in Figure \ref{fig:Ne_Fe-O_Fe} \red{because} the Ne yields are not shown in \citet{Umeda2008}.

\red{Since} SNeIa show lower O/Fe than BrHNe and PISNe, combination of SNeIa and CCSNe could explain the low O/Fe value. Here, the timescale of these supernovae is important. In Figure \ref{fig:z-O_Fe}, we present a redshift evolution of O/Fe for galaxies since $z=20$ ($\sim200$ Myr after the Big Bang), when the first star formation is expected to occur from the standard $\Lambda$-CDM scenario (e.g., \citealt{Tegmark1997}). While BrHNe and PISNe instantaneously occur after the formation of massive stars, SNeIa need the delay time due to \red{longer lifetime of the progenitor} of white dwarfs and gas accretion to occur. Because the typical delay time of SNeIa is $\sim10^{8.5-9}$ yr (e.g., \citealt{Chen2021}), SNeIa have difficulty in causing Fe enrichment in GN-z11 at as early as $z=10.60$, when the age of the universe is only $430$ Myr. However, considering that the first SNIa has a delay time of $\sim30$ Myr (e.g., \citealt{Maiolino2019}), the short delay time and early star formation might accomplish the low O/Fe ratio. In conclusion, we suggest that the Fe enrichment of GN-z11 at $z=10.60$ is caused by BrHNe and/or PISNe, or that the short delay time and early star formation allow the Fe enrichment of SNeIa.

\subsection{Connection between GN-z11 and GC Formation} \label{subsec:GC}
\begin{figure}[t]
    \centering
    \includegraphics[scale=0.45]{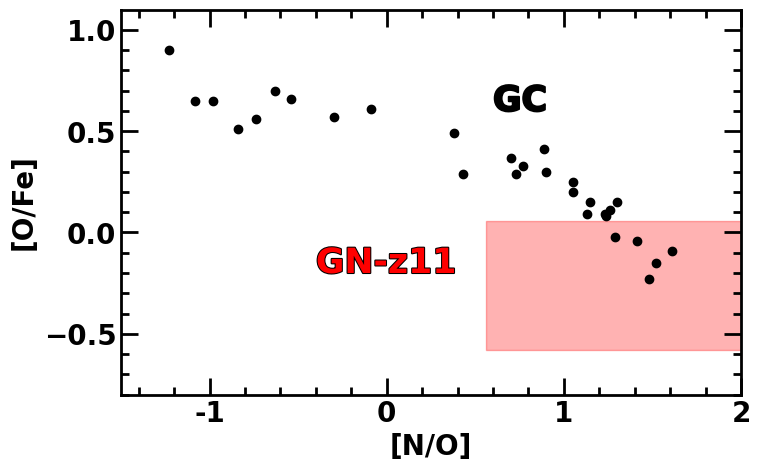}
    \caption{Comparison of \red{O/Fe} vs. N/O for GN-z11 and GC stars. The red-shaded region denotes the abundance measurements of GN-z11 in this work and \citet{Cameron2023}. The black circles represent the GC stars \citep{Melendez2003,Carretta2005,Yong2005,Yong2008,Pasquini2008}.}
    \label{fig:N_O-O_Fe}
\end{figure}
The high N/O value ($\mathrm{[N/O]}>0.61$) of GN-z11 reported by \citet{Cameron2023} is comparable to those of some GC stars \citep{Carretta2005}, which \red{suggests that GN-z11 may be the site of GC formation}, as discussed in recent studies (e.g., \citealt{Isobe2023b,Senchyna2024}). The connection \red{between GN-z11 and GC formation} is also \red{suggested} by the low C/N value of GN-z11 and GC stars, which can be originated from CNO-cycled gas \citep{Isobe2023b}. While nitrogen is produced by CNO-cycle in outer H-burning layer of stars and ejected via the stellar winds, iron is produced in the core of stars and ejected by the supernovae. There appears to be a contradiction between the high N/O and low O/Fe values of GN-z11. However, we find that some GC stars \citep{Melendez2003,Carretta2005} also have abundance ratios of high N/O and low O/Fe which are comparable to those of GN-z11, as shown in Figure \ref{fig:N_O-O_Fe}. We discuss possible origins of such unusual abundance ratios in the next paragraph. The O/Fe of GN-z11 is low compared to the MW stars and lower redshift galaxies, but not too low to rule out the connection between GN-z11 and GC formation. Instead, the O/Fe measurement of GN-z11 \red{supports} the connection with GC stars with high N/O. \red{We note that the GC stars which have similar N/O and O/Fe ratios to those of GN-z11 show the low Fe/H and O/H ratios compared to GN-z11 (see Figures \ref{fig:O_H-O_Fe} and \ref{fig:Fe_H-O_Fe}). The apparent deviation of Fe/H and O/H values can be explained by metal dilution caused by primordial (i.e., only hydrogen) gas inflow (e.g., \citealt{Kojima2021,Isobe2022}).}

We discuss a possible scenario to explain the abundance ratios of low O/Fe and high N/O found in GN-z11 and GC stars. In the early universe, metal-poor gas suppresses the gas cooling and makes the Jeans mass higher, which results in the formation of many massive stars like Population III stars (e.g., \citealt{Hirano2014,Hirano2015}).  If such first generations of massive stars cause BrHNe and/or PISNe, the Fe-rich gas is ejected due to the high explosion energy (Alternatively, if the delay time is short, the Fe enrichment can be accomplished by SNeIa). The second generation of stars then form from part of the Fe-rich gas and eject N-rich gas via the stellar wind, which results in the low O/Fe and high N/O ratios. To confirm whether this scenario can actually explain the observed low O/Fe and high N/O values, we need to quantitatively test the scenario with theoretical models.
 
%\clearpage
\section{Summary} \label{sec:summary}
In this paper, we present our measurements of the O/Fe value of GN-z11 at $z=10.60$. We measure the stellar metallicity by fitting the model spectra constructed from the stellar population synthesis models and nebular continua to the stacked spectra of GN-z11 obtained from the JWST/NIRSpec observations conducted in the JADES program. We carefully evaluate the inferred stellar metallicity, comparing the fitting results from both the grating and prism spectra, and results of the independent method using the 1978 index \citep{Rix2004}. Our major findings are summarized below:
\begin{itemize}
    \item[1.] We find a low O/Fe value of $\mathrm{[O/Fe]}=\red{-0.37^{+0.43}_{-0.22}}$ for GN-z11 compared to the MW stars and lower redshift galaxies at $z\sim2-3$.    Considering the typical delay time for white dwarfs and gas accretion, it is difficult to explain Fe enrichment at as early as $z=10.60$ (only $\sim430$ Myr after the Big Bang) by SNeIa. The abundance ratios of O/Fe and Ne/Fe of GN-z11 are comparable to the yields of BrHNe and PISNe. We conclude that Fe can be enhanced in GN-z11 by the combination of short delay time and early star formation, or occurrence of BrHNe and/or PISNe.
    \item[2.] We compare abundance ratios of O/Fe and N/O for GN-z11 and GC stars, of which connection is suggested by similarly high N/O ratio in recent studies. We find that O/Fe of GN-z11 is not too low to rule out the connection between GN-z11 and GC formation, but rather confirm the connection if metal dilution is caused by primordial gas inflow.
\end{itemize}

%% Also note that the akcnowlodgment environment does not support long amounts of text. If you have a lot of people and institutions to acknowledge, do not use this command. Instead, create a new \section{Acknowledgments}.

\section*{Acknowledgements} 
We thank Hajime Fukushima, \red{Keita Fukushima, Daisuke Toyouchi,} Yi Xu, and Hidenobu Yajima for the valuable discussions on this work. This work is based on observations made with the NASA/ESA/CSA James Webb Space Telescope. The data were obtained from the Mikulski Archive for Space Telescopes at the Space Telescope Science Institute, which is operated by the Association of Universities for Research in Astronomy, Inc., under NASA contract NAS 5-03127 for JWST. The specific observations analyzed can be accessed via \dataset[10.17909/8tdj-8n28]{https://doi.org/10.17909/8tdj-8n28}. These observations are associated with programs GTO-1181 (JADES). We thank JADES team for publicly releasing reduced spectra and catalog from the JADES survey. This publication is based upon work supported by the World Premier International Research Center Initiative (WPI Initiative), MEXT, Japan, and KAKENHI (20H00180, 21H04467) through Japan Society for the Promotion of Science. K.N. has been supported by KAKENHI Grant Numbers JP21H04499 and JP23K03452. This work has been supported by JSPS KAKENHI Grant Numbers JP24KJ0202 (Y.I.) and JP21K13956 (\red{D.K.}). M.N. was supported by JST, the establishment of university fellowships towards the creation of science technology innovation, Grant Number JPMJFS2136. This work was supported by the joint research program of the Institute for Cosmic Ray Research (ICRR), the University of Tokyo.
%Ken'ichi Nomoto has been supported by KAKENHI Grant Numbers JP21H04499 and JP23K03452. Yuki Isobe has been supported by KAKENHI Grant Number 24KJ0202.


%% citation or url to the code inside ()s when available

\software{NumPy \citep{Harris2020}, matplotlib \citep{Hunter2007}, SciPy \citep{Virtanen2020}, Astropy \citep{Astropy2013,Astropy2018,Astropy2022}, emcee \citep{Foreman2013}, BPASS v2.0 \citep{Eldridge&Stanway2016,Stanway2016}, and \textsc{Cloudy} v23.01 \citep{Ferland1998,Gunasekera2023}}.





%\clearpage
\bibliographystyle{aasjournal}
\bibliography{library.bib}



\clearpage
\if0
\appendix
\restartappendixnumbering
%\section{Comparison of Lyman alpha equivalent width measured with the grating and prism spectra}
\label{sec:apA}
\fi

\end{document}